\documentclass[12pt]{article}

\def\tcon{\tilde{t}}
\def\tbmu{\tilde{B}_{\tilde{\mu}}}
\def\tmu{\tilde{\mu}}
\usepackage{graphicx}
\input epsf

\begin{document}

\begin{titlepage}
\begin{flushright}
CALT-68-2545\\
ITEP-TH-30/05
\end{flushright}

\begin{center}
{\Large\bf $ $ \\ $ $ \\
An action variable of the sine-Gordon model}\\
\bigskip\bigskip\bigskip
{\large Andrei Mikhailov}
\\
\bigskip\bigskip
{\it California Institute of Technology 452-48,
Pasadena CA 91125 \\
\bigskip
and\\
\bigskip
Institute for Theoretical and 
Experimental Physics, \\
117259, Bol. Cheremushkinskaya, 25, 
Moscow, Russia}\\

\vskip 1cm
\end{center}

\begin{abstract}
	It was conjectured that the classical bosonic string
	in AdS times a sphere has a special action variable which corresponds
	to the length of the operator on the field theory side.
	We discuss the analogous action variable in the sine-Gordon
	model. We explain the relation between this action variable
	and the B\"acklund transformations and show that the corresponding
	hidden symmetry acts on breathers by shifting their phase. 
	It can be considered a nonlinear analogue of splitting
	the solution of the free field equations into the positive-
	and negative-frequency part.
\end{abstract}

\end{titlepage}
\section{Introduction.}
Studies of classical strings in $AdS_5\times S^5$ was an important
part of the recent work on the AdS/CFT correspondence.
It was observed that the energies of the fast moving classical
strings reproduce the anomalous dimension of the 
field theory operators with the large R-charge, at least
in the first and probably the second order of the perturbation
theory  
\cite{FT02,Tseytlin,Russo,MinahanZarembo,FT03,FTQ,Kruczenski,KRT}.

Classical superstring in $AdS_5\times S^5$ is an integrable system.
An important tool in the study of this system is the super-Yangian symmetry 
discussed in \cite{MandalSuryanarayanaWadia,BPR}.
The nonabelian dressing symmetries were also found in the classical
Yang-Mills theory, see the recent discussion in \cite{MartinWolf} 
and the references therein.
It was conjectured in 
\cite{DNW1,DNW2,DolanNappi} that the super-Yangian symmetry is
also a symmetry of the Yang-Mills perturbation theory. It was shown
that the one-loop anomalous dimension is proportional
to the first Casimir operator in the Yangian representation. 
It is natural to conjecture that the higher loop contributions
to the anomalous dimension correspond to the higher Casimirs in
the Yangian representation, in the following sense. 
We conjecture that there are infinitely many operators
$C_1, C_2, C_3, \ldots$ acting on the spin chain Hilbert
space, commuting:
\[
[C_i,C_j]=0
\]
and depending on the 'tHooft coupling constant $\lambda$ as power series:
\[
C_j=C_{j,0}+\lambda C_{j,1}+\lambda^2 C_{j,2}+\ldots
\]
and such that the anomalous dimension $\Delta$ has an expansion:
\begin{equation}\label{DeltaInFieldTheory}
\Delta=\lambda C_1 +\lambda^2 C_2+\lambda^3 C_3 +\ldots
\end{equation}
It should be true that $C_{j,0}$ involves $j$ nearest neighbors in the chain.
Notice that $C_j$ explicitly depends on $\lambda$.
Therefore the expansion of the anomalous dimension is:
\[
	\Delta=\lambda \Delta_1+\lambda^2 \Delta_2 +\lambda^3\Delta_3+\ldots
\]
where
\[
\Delta_j=\sum_{k=0}^{j-1} C_{j-k,k}
\]
Notice that $[\Delta_j,\Delta_k]\neq 0$ for $j\neq k$.

We have argued in \cite{Anomalous} that the 
relation analogous to (\ref{DeltaInFieldTheory})
holds for the classical string in $AdS_5\times S^5$, 
the conserved charge $\lambda^jC_j$ corresponding to the
$j$-th improved Pohlmeyer charge.
We suggested to identify
the anomalous dimension with the deck transformation acting 
on the phase space of the classical string in AdS times a sphere.
The deck transformation can be defined as the
action of the center of the conformal group. It is a {\em geometric} symmetry 
of the classical string; it comes from 
a geometric symmetry of the AdS space. 
But it can be
expressed in terms of the {\em hidden} symmetries. String theory
in AdS times a sphere has an infinite family of local conserved
charges, the Pohlmeyer charges \cite{Pohlmeyer}. 
These Pohlmeyer charges can be thought of as the classical
limit of the Yangian Casimirs. We have argued in \cite{Anomalous}
that the deck transformation is in fact generated by an infinite
linear combination of the Pohlmeyer charges; the coefficients
of this linear combination were fixed in \cite{PWL}.
We used in our arguments the existence of a special
action variable in the theory of the classical string on $S^n$
which was discussed in \cite{Notes} following
\cite{ArutyunovStaudacher,Engquist,KT}.
The special property
of this particular action variable is that in each order
of the null-surface perturbation theory 
\cite{dVGN,SpeedingStrings}
it is given by a local 
expression\footnote{The results of 
\cite{ArutyunovFrolov0411,BKSZ,AAT} imply that this 
property of the action variable does not hold for the
superstring. Indeed, the construction of the action variable
in Section 4 of \cite{Notes} used the fact
that the classical bosonic sigma-model splits into the $AdS_5$ part
and the $S^5$ part. But the fermions ``glue together'' the
AdS and the sphere. It was argued in \cite{ArutyunovFrolov0411,BKSZ,AAT} 
that this action variable still exists in the supersymmetric case 
but is not local. What survives the supersymmetric extension
is the statement that the deck transformation 
in each order of the null-surface perturbation theory is
generated by a finite sum of the local conserved charges
(the classical analogues of the super-Yangian Casimirs).
Indeed, the definition of the deck transformation
does not require the splitting of the sigma-model into two parts
and the locality of the deck transformation in the perturbation
theory is manifest; it is essentially a consequence of
the worldsheet causality. I want to thank N.~Beisert for
a discussion of this subject.}. In each
order of the perturbation theory we can approximate
this action variable by a finite sum of the Pohlmeyer 
charges.
This action variable corresponds to the {\em length} of the
spin chain on the field theory side \cite{Minahan}.

This ``length'' was studied for the finite gap solutions
in \cite{BKS,BKSZ}.
Here we will study it for the rational solutions.
We will consider  the
simplest case of the classical string on ${\bf R}\times S^2$.
This system is related \cite{Pohlmeyer} to the sine-Gordon model and we
will actually discuss mostly the sine-Gordon model. 
The existence of the special action variable can be understood
locally on the worldsheet, at least in the null-surface perturbation theory. 
Therefore to study this action variable
we do not have to impose the periodicity conditions on the spacial
direction of the string worldsheet; we can formally consider infinitely
long strings. This allows us to use the rational solutions of the
sine-Gordon equation which are probably ``simpler'' than the
finite gap solutions studied in 
\cite{KMMZ,KazakovZarembo,BKS,SchaferNameki,BKSZ}
(at least if we consider the elementary functions 
``simpler'' than the theta functions.)

In Section \ref{sec:SineGordon} we discuss the relation between
the classical string propagating on ${\bf R}\times S^2$ and
the sine-Gordon model. In Section \ref{sec:Tau} we 
discuss the tau-function and bilinear identities. 
In Section \ref{sec:Backlund} we discuss B\"acklund transformations
and define the ``hidden'' symmetry $U(1)_L$. 
In Section \ref{sec:FreeFieldLimit} we consider
the plane wave limit. In Section \ref{sec:Breather} we explain 
how $U(1)_L$ acts on breathers.
In Section \ref{sec:Improved} we discuss the ``improved'' currents
and show that $U(1)_L$ has a local expansion in the null-surface
perturbation theory.
In Section \ref{sec:Summary} we summarize our construction
of the action variable
and outline the analogous construction for the $O(N)$ 
sigma model.

\section{Sine-Gordon and string on ${\bf R}\times S^2$.}\label{sec:SineGordon}
\subsection{Sine-Gordon
equation from the classical string.}
The sine-Gordon model is one of the simplest exactly
solvable models of interacting relativistic fields,
and the bosonic string propagating on
${\bf R}\times S^2$ is one of 
the simplest nonlinear string worldsheet theories.
On the level of classical equations of motion these
two models are equivalent.

Consider the classical string propagating on ${\bf R}\times S^2$.
Let $t$ denote the time coordinate parametrizing $\bf R$.
We will choose the conformal coordinates $(\tau,\sigma)$ on the
worldsheet so that the induced metric is proportional to 
$d\tau^2-d\sigma^2$. We will also fix the residual freedom in 
the choice of the conformal coordinates by putting $\tau=t$.
We can parametrize the sphere by unit vectors $\vec{n}$; 
the embedding of the classical string in $S^2$ is parametrized
by $\vec{n}(\tau,\sigma)$.
The worldsheet equations of motion are:
\begin{equation}
	(\partial_{\tau}^2-\partial_{\sigma}^2)\vec{n}=
	-[ (\partial_{\tau}\vec{n})^2 - (\partial_{\sigma}\vec{n})^2 ]
	\vec{n}
\end{equation}
These equations of motion follow from the constraints:
\begin{eqnarray}\label{VirasoroConstraints}
	\left({\partial\vec{n}\over\partial\tau}\right)^2+
	\left({\partial\vec{n}\over\partial\sigma}\right)^2=1
	\\[5pt]
	\left({\partial\vec{n}\over\partial\tau},
	      {\partial\vec{n}\over\partial\sigma}\right)=0
\end{eqnarray}
The map to the sine-Gordon model is given by \cite{Pohlmeyer}:
\begin{equation}\label{FromStringToSG}
	\cos 2\phi=\left({\partial\vec{n}\over\partial\tau}\right)^2
	-\left({\partial\vec{n}\over\partial\sigma}\right)^2
\end{equation}
In other words
\begin{equation}\label{WithAbsValues}
	\left|\partial_{\tau}\vec{n}\right|=
	|\cos\phi|,\;\;\;\;\;
	\left|\partial_{\sigma}\vec{n}\right|=
	|\sin\phi|
\end{equation}
The Virasoro constraints (\ref{VirasoroConstraints}) are equivalent to
the sine-Gordon equation:
\begin{equation}
	\left[\partial^2_{\tau}-\partial^2_{\sigma}\right]\phi=
	-{1\over 2}\sin 2\phi
\end{equation}
What can we say about the inverse map, from $\phi$ to $\vec{n}$?
Let us consider the limit when the string moves very fast.

\subsection{Null-surface limit, plane wave limit, free field limit.}
When the string moves very fast
$|\partial_{\tau}\vec{n}|>>|\partial_{\sigma}\vec{n}|$.
As in \cite{dVGN,SpeedingStrings} 
we replace $\sigma$ with $s=\epsilon\sigma$
where $\epsilon$ is a small parameter. Because of (\ref{WithAbsValues}) 
we should also replace
$\phi$ with $\epsilon\psi$; the new field $\psi(\tau,s)$ will be finite
in the null-surface limit:
\begin{equation}\label{NSL}
	\sigma=\epsilon^{-1} s,\;\;\;\;\;
	\phi(\tau,\sigma)=\epsilon\psi(\tau,s)
\end{equation}
The string embedding $\vec{n}$ satisfies:
\begin{eqnarray}
&&	|\partial_s\vec{n}|=\psi-{\epsilon^2\over 6}\psi^3+\ldots
	\label{Dsn}
	\\
&&	|\partial_{\tau}\vec{n}|=1-{\epsilon^2\over 2}\psi^2+\ldots
\end{eqnarray}
The rescaled sine-Gordon field $\psi$ satisfies:
\begin{equation}\label{RescaledSG}
	[\partial_{\tau}^2-\epsilon^2\partial_s^2]\psi=
	-\psi+{\epsilon^2\over 6}\psi^3+\ldots
\end{equation}
In the strict null-surface limit $\epsilon=0$ and
\begin{equation}\label{VariationOfGeodesic}
\psi(\tau,s)=a(s)\cos(\tau+\alpha(s)) 
\end{equation}
In this limit the string worldsheet is a collection of null-geodesics.
The $S^2$-part is therefore a collection of equators of $S^2$.
For each point $(\tau_0,s_0)$ on the worldsheet the intersection
of $S^2$ with the 2-plane generated by $\vec{n}(\tau_0,s_0)$ and 
$\partial_{\tau}\vec{n}(\tau_0,s_0)$ is the corresponding equator;
this equator can be parametrized by the vector 
$\vec{V}=[\vec{n}\times \partial_{\tau}\vec{n}]$.
We have
$$\vec{n}(\tau,s_0)=\cos(\tau-\tau_0) \vec{n}(\tau_0,s_0) 
+\sin(\tau-\tau_0) \partial_{\tau_0}\vec{n}(\tau_0,s_0)$$
Eqs. (\ref{Dsn}) and (\ref{VariationOfGeodesic}) show that the one-parameter
family of equators forming the null-surface is given by the equation
$$
	\partial_s\vec{V}(s)=
	[(-a(s)\; \vec{n}\;\cos(\tau_0+\alpha(s))+a(s)\; 
	\partial_{\tau}\vec{n}\;\sin(\tau_0+\alpha(s)))
	\times \vec{V}(s)]
$$
where $\alpha(s)$ and $a(s)$ are determined from $\psi(\tau,s)$ by
(\ref{VariationOfGeodesic}).
Therefore in the limit $\epsilon\to 0$ the null-surface is determined
by $\lim_{\epsilon\to 0} \epsilon^{-1}\phi$.
It should be possible in principle to extend this analysis to higher orders 
and find
the extremal surface corresponding to the solution $\phi$ of the
sine-Gordon equation. The extremal surface is determined by $\phi$
up to the rotations of $S^2$. 

Another important limit is the {\em plane wave limit.}
To get to the plane wave limit we first go to the null-surface limit
(\ref{NSL}) and then take an additional rescaling $\psi=\epsilon_1 \chi$.
In the strict limit $\epsilon_1=0$ the equations for $\chi$ become
linear:
\begin{equation}
[\partial_{\tau}^2-\epsilon^2\partial_s^2]\chi=-\chi
\end{equation}
The plane wave limit is therefore the free field limit.

\subsection{Poisson structure.}
The canonical Poisson structure of the classical string
on ${\bf R}\times S^2$ does not agree with the canonical
Poisson structure of the sine-Gordon model. But it corresponds
to another Poisson structure of the sine-Gordon model,
which is compatible with the canonical one \cite{Nonlocal}.
Therefore, the Hamiltonian flows generated by the local
conserved charges of the sine-Gordon model
should differ from the flows of the Pohlmeyer charges
of the classical string
only by the relabelling of the charges. 
This means that the action variable of the sine-Gordon 
model which we discuss in this paper corresponds to the
action variable of the classical string, which is also
generated by an infinite linear combination of the local 
Pohlmeyer charges. 

\section{Rational solutions.}\label{sec:Tau}
\subsection{Tau-functions and the dependence on higher times.}
In this section we will discuss the dependence of the sine-Gordon
solutions on the ``higher times'' following mostly 
\cite{BabelonBernardAffine,FaddeevTakhtajan,KMMMZ}.
We will first introduce the tau-functions and then explain how they
are related to the solutions of the sine-Gordon equations.

The tau-functions for the rational solutions are
\begin{eqnarray}
	\tau_{\pm}&=&\det (1\pm {\cal V}) \\[5pt]
	{\cal V}_{jk}&=& 
	2ib_jb_k{\sqrt{\lambda_j\lambda_k}\over \lambda_j+\lambda_k}
	\times\label{Tau}\\
&&	\times\exp\left[
	\sum_p t_{2p+1}(\lambda_j^{2p+1}+\lambda_k^{2p+1})
	-\sum_p \tcon_{2p+1}(\lambda_j^{-2p-1}+\lambda_k^{-2p-1})
	\right]
	\nonumber
\end{eqnarray}
Here $b_j$ and $\lambda_j$, $j=1,\ldots,N$ are parameters characterizing
the solution, and $t_{2p+1},\tilde{t}_{2p+1}$, $p=0,1,2,\ldots$ are
the so-called times. We identify $t_1={1\over 4}(\tau+\sigma)$
and $\tilde{t}_1={1\over 4}(\tau-\sigma)$. The ``higher'' times
$t_3,\tilde{t}_3,t_5,\tilde{t}_5,\ldots$ correspond to the higher
conserved charges. Changing the higher times corresponds to the
motion on the ``Liouville torus'' in the phase space. 
Rational solutions correspond to finite $N$; the tau-functions of the
rational solutions are the determinants of  the $N\times N$ matrices
$\delta_{ij}\pm {\cal V}_{ij}$.

Let us consider the left and right B\"acklund transformations:
\begin{eqnarray}\label{BacklundShift}
&&	B_{\mu}.\tau_{\pm}(\{t_{2p+1}\},\{\tcon_{2q+1}\})=
	\tau_{\pm}\left(\left\{t_{2p+1}-{\mu^{-2p-1}\over 2p+1}\right\},
	\left\{\tcon_{2q+1}\right\}\right)
	\\[5pt]
&&	\tilde{B}_{\tilde{\mu}}.\tau_{\pm}(\{t_{2p+1}\},\{\tcon_{2q+1}\})=
	\tau_{\pm}\left(\left\{t_{2p+1}\right\},
	\left\{\tcon_{2q+1}+{\tilde{\mu}^{2q+1}\over 2q+1}\right\}\right)
	\label{BacklundShiftRight}
\end{eqnarray}
where $\mu$ and $\tilde{\mu}$ are constant parameters.
The tau-functions satisfy the following bilinear identities:
\begin{eqnarray}
&&	B_{\mu}B_{\nu}\tau_+\tau_-+
	\tau_+ B_{\mu}B_{\nu}\tau_-=
	B_{\mu}\tau_+ B_{\nu}\tau_-+
	B_{\nu}\tau_+ B_{\mu}\tau_-
	\nonumber
	\\[8pt]
&&	{\nu-\mu\over\nu+\mu}
	(B_{\mu}B_{\nu}\tau_+\tau_--
	\tau_+ B_{\mu}B_{\nu}\tau_-)=
	\label{LeftBilinearIdentities}\\
&&	=B_{\mu}\tau_+ B_{\nu}\tau_--
	B_{\nu}\tau_+ B_{\mu}\tau_-\nonumber
\end{eqnarray}
\begin{eqnarray}
&&	\tilde{B}_{\tilde{\mu}}
	\tilde{B}_{\tilde{\nu}}\tau_+\tau_-+
	\tau_+ \tilde{B}_{\tilde{\mu}}
	\tilde{B}_{\tilde{\nu}}\tau_-=
	\tilde{B}_{\tilde{\mu}}\tau_+ 
	\tilde{B}_{\tilde{\nu}}\tau_-+
	\tilde{B}_{\tilde{\nu}}\tau_+ 
	\tilde{B}_{\tilde{\mu}}\tau_-
	\nonumber
	\\[8pt]
	&&	{\tilde{\nu}-\tilde{\mu}\over\tilde{\nu}+\tilde{\mu}}
	(\tilde{B}_{\tilde{\mu}}\tilde{B}_{\tilde{\nu}}\tau_+
	\tau_--\tau_+ 
	\tilde{B}_{\tilde{\mu}}\tilde{B}_{\tilde{\nu}}\tau_-)=
	\label{RightBilinearIdentities}\\
&&	=-\tilde{B}_{\tilde{\mu}}\tau_+ 
	\tilde{B}_{\tilde{\nu}}\tau_-+
	\tilde{B}_{\tilde{\nu}}\tau_+ 
	\tilde{B}_{\tilde{\mu}}\tau_-\nonumber
\end{eqnarray}
\begin{eqnarray}
&&	B_{\mu}\tilde{B}_{\tilde{\nu}}\tau_+\tau_++
	B_{\mu}\tilde{B}_{\tilde{\nu}}\tau_-\tau_-=
	B_{\mu}\tau_+\tilde{B}_{\tilde{\nu}}\tau_++
	B_{\mu}\tau_-\tilde{B}_{\tilde{\nu}}\tau_-
	\nonumber
	\\[8pt]
&&	{\tilde{\nu}-\mu\over\tilde{\nu}+\mu}
	(B_{\mu}\tilde{B}_{\tilde{\nu}}\tau_-
	\tau_--\tau_+ 
	B_{\mu}\tilde{B}_{\tilde{\nu}}\tau_+)=
	\label{Mixed} \\
&&	=B_{\mu}\tau_+ 
	\tilde{B}_{\tilde{\nu}}\tau_+-
	\tilde{B}_{\tilde{\nu}}\tau_- 
	B_{\mu}\tau_-\nonumber
\end{eqnarray}
These bilinear identities can be derived from the free fermion representation
of the tau-function as explained for example is \cite{KMMMZ}.
We introduce free fermions $\psi(\mu)=\sum_{m\in{\bf Z}} \psi_m\mu^{m-1/2}$
and $\tilde{\psi}(\mu)=\sum_{m\in{\bf Z}}\tilde{\psi}_m\mu^{-m+1/2}$,
$\{\psi_m,\tilde{\psi}_n\}=\delta_{mn}$. The ``vacuum vectors''
are labeled by $k\in{\bf Z}$ so that 
$\langle k| \psi(\lambda_1)\tilde{\psi}(\lambda_2) | k\rangle =
\left({\lambda_1\over\lambda_2}\right)^k {\sqrt{\lambda_1\lambda_2}\over
\lambda_1-\lambda_2}$ for $|\lambda_1|>|\lambda_2|$.
Let us put $k_+=0$ and $k_-=1$. We have
\begin{eqnarray}
&&	\tau_{\pm} = 
e^{-\sum (2p+1)t_{2p+1}\tilde{t}_{2p+1}}
	\times  \\
&&	\times \langle k_{\pm} | 
	e^{\sum t_{2p+1}\psi_n\tilde{\psi}_{n+2p+1}}
	\prod_{j=1}^N\left[1+2b_j^2\psi(\lambda_j)\tilde{\psi}(-\lambda_j)
	\right]
	e^{\sum \tilde{t}_{2p+1}\psi_n\tilde{\psi}_{n-2p-1}}
	|k_{\pm}\rangle
	\nonumber
\end{eqnarray}
Eq. (\ref{LeftBilinearIdentities}) is Eq. (2.42) of \cite{KMMMZ} 
if we take into account that 
\begin{equation}
	\tau_-=\lim_{\mu\to 0}B_{\mu}\tau_+
\end{equation}
Let us study some differential equations following from the 
bilinear identities.
From (\ref{Mixed}) we have at the first order in $\tilde{\nu}/\mu$:
\begin{eqnarray}
	\tau_-\partial_{t_1}\partial_{\tilde{t}_1}\tau_- -
	\partial_{t_1}\tau_- \partial_{\tilde{t}_1}\tau_- =
	-\tau_-^2+\tau_+^2
	\\
	\tau_+\partial_{t_1}\partial_{\tilde{t}_1}\tau_+ -
	\partial_{t_1}\tau_+ \partial_{\tilde{t}_1}\tau_+ =
	-\tau_+^2+\tau_-^2
\end{eqnarray}
Therefore the equations of motion for the sine-Gordon model
\begin{equation}\label{SineGordon}
	{\partial^2\over\partial t_1 \partial \tilde{t}_1}
	\phi = -2\sin 2\phi
\end{equation}
follow if we set 
\begin{equation}\label{PhiFromTau}
	\phi=i\log{\tau_+\over\tau_-}
\end{equation}
Expanding (\ref{LeftBilinearIdentities}) in the powers of $1\over \nu$
we have
\begin{equation}
	\partial_{t_1}\tau_- B_{\mu}\tau_+-
	\tau_- \partial_{t_1}B_{\mu}\tau_+=
	\mu(\tau_-B_{\mu}\tau_+-\tau_+ B_{\mu}\tau_-)
\end{equation}
and the same equation with $\tau_+$ and $\tau_-$ exchanged.
This can be rewritten as the first order differential equation
relating $B_{\mu}\phi$ to $\phi$:
\begin{equation}\label{LeftBacklundSmall}
	{\partial\over\partial t_1}(B_{\mu}\phi+\phi)
	=-2\mu\sin(B_{\mu}\phi-\phi)
\end{equation}
Expanding (\ref{Mixed}) in powers of $\tilde{\nu}$ we get:
\begin{equation}
	\tau_+\partial_{\tilde{t}_1}B_{\mu}\tau_+-
	\partial_{\tilde{t}_1}\tau_+ B_{\mu}\tau_+ =
	{1\over \mu} (\tau_+B_{\mu}\tau_+ -\tau_- B_{\mu}\tau_-)
\end{equation}
and the same equation with $\tau_+$ exchanged with $\tau_-$.
This gives us the second equation relating $B_{\mu}\phi$ to $\phi$:
\begin{equation}\label{LeftBacklundLarge}
	{\partial\over\partial \tilde{t}_1}
	(B_{\mu}\phi-\phi)=
	{2\over\mu}\sin(B_{\mu}\phi+\phi)
\end{equation}
Expanding (\ref{RightBilinearIdentities}) in the powers of 
$\tilde{\nu}$ we get 
\begin{equation}
	\tau_-\partial_{\tilde{t}_1}\tilde{B}_{\tilde{\mu}}\tau_+  -
	\tilde{B}_{\tilde{\mu}}\tau_+ \partial_{\tilde{t}_1} \tau_-=
	{1\over \tilde{\mu}}(\tau_-\tilde{B}_{\tilde{\mu}}\tau_+ -
	\tau_+\tilde{B}_{\tilde{\mu}}\tau_-)
\end{equation}
and the same equation with $\tau_+$ and $\tau_-$ exchanged.
This gives us the equation relating $\tilde{B}_{\tilde{\mu}}\phi$ 
to $\phi$:
\begin{equation}\label{RightBacklundSmall}
	{\partial\over\partial \tilde{t}_1}
	(\tilde{B}_{\tilde{\mu}}\phi+\phi)=
	{2\over\tilde{\mu}}\sin(\tilde{B}_{\tilde{\mu}}\phi-\phi)
\end{equation}
The second equation follows from (\ref{Mixed}):
\begin{equation}\label{RightBacklundLarge}
	{\partial\over\partial t_1}
	(\tilde{B}_{\tilde{\mu}}\phi-\phi)=
	-2\tilde{\mu}\sin(\tilde{B}_{\tilde{\mu}}\phi+\phi)
\end{equation}
Equations (\ref{LeftBacklundSmall}), (\ref{LeftBacklundLarge}),
(\ref{RightBacklundSmall}) and (\ref{RightBacklundLarge}) are usually
taken as the definition of the left and right B\"acklund transformations.
These equations do not determine $B_{\mu}\phi$ and
$\tilde{B}_{\tilde{\mu}}\phi$ unambiguously from $\phi$ because
there are integration constants.  
Eqs. (\ref{BacklundShift}) and (\ref{BacklundShiftRight}) provide
a particular solution. 

The B\"acklund transformations for the sine-Gordon field 
correspond to the B\"acklund transformations for the classical string.
If $\vec{n}(\tau,\sigma)$ is a string worldsheet and
$\phi$ is  the corresponding solution of the sine-Gordon model defined by 
Eq. (\ref{FromStringToSG}) then
\begin{equation}
	B_{\mu}\vec{n}=
	{1-\mu^{-2}\over 1+\mu^{-2}}\vec{n}-
	{\mu^{-1}\over 1+\mu^{-2}}
	\left({\sin(\phi-B_{\mu}\phi)\over \sin(2\phi)}
	\partial_{\tilde{t}_1}\vec{n}+
	{\sin(\phi+B_{\mu}\phi)\over \sin(2\phi)}
	\partial_{t_1}\vec{n}\right)
\end{equation}
satisfies
\begin{eqnarray}\label{LeftBacklundVector}
&&	\partial_{\tilde{t}_1}(B_{\mu}\vec{n}-\vec{n})=
	-{1\over 2}(1+\mu^{-2})(B_{\mu}\vec{n},\partial_{\tilde{t}_1}\vec{n})
	(B_{\mu}\vec{n}+\vec{n})
	\nonumber
	\\[-7pt]
	\\[-7pt]
&&	\partial_{t_1}(B_{\mu}\vec{n}+\vec{n})=
	{1\over 2}(1+\mu^2)(B_{\mu}\vec{n},\partial_{t_1}\vec{n})
	(B_{\mu}\vec{n}-\vec{n})
	\nonumber
\end{eqnarray}
and
\begin{equation}
	\tbmu \vec{n}=
	{1-\tmu^2\over 1+\tmu^2}\vec{n}
	+{\tmu\over 1+\tmu^2}\left(
	{\sin(\phi-\tbmu\phi)\over \sin(2\phi)}\partial_{t_1}\vec{n}
	+
	{\sin(\phi+\tbmu\phi)\over \sin (2\phi)}
	\partial_{\tilde{t}_1}\vec{n}\right)
\end{equation}
satisfies
\begin{eqnarray}\label{RightBacklundVector}
&&	\partial_{t_1}(\tbmu \vec{n} -\vec{n})=
	-{1\over 2}(1+\tmu^2)(\tbmu\vec{n},\partial_{t_1}\vec{n})
	(\tbmu\vec{n}+\vec{n})
	\nonumber
	\\[-7pt]
	\\[-7pt]
&&	\partial_{\tilde{t}_1}(\tbmu\vec{n}+\vec{n})=
	{1\over 2}(1+\tmu^{-2})(\tbmu\vec{n},\partial_{\tilde{t}_1}\vec{n})
	(\tbmu\vec{n}-\vec{n})
	\nonumber
\end{eqnarray}
The relation between $B_{\mu}\vec{n}$ and $B_{\mu}\phi$,
and between $\tbmu\vec{n}$ and $\tbmu\phi$, is given by 
Eq. (\ref{FromStringToSG}).
The relation between B\"acklund transformations in $O(3)$ model and 
sine-Gordon model has been previously discussed in \cite{NeveuPapanicolaou}.

\subsection{The reality conditions and a restriction on
the class of solutions.}
To get the real solutions of the sine-Gordon theory we need
$\tau_+$ to be the complex conjugate of $\tau_-$. 
This can be achieved if
the parameters $\lambda_i$ come
in pairs $\lambda_k$ and $\lambda_{N-k}$ 
such that $\lambda_k=\overline{\lambda}_{N-k}$ and 
$b_k=\overline{b}_{N-k}$.
We want to restrict ourselves with considering only the solutions
for which all $\lambda_j$  have a nonzero imaginary part:
\begin{equation}
	\mbox{Im}\;\lambda_j\neq 0
\end{equation}
The purely real $\lambda_j$ would lead to kinks; 
we consider the solutions
with kinks too far from being the fast moving strings.

General solutions of the sine-Gordon equations on a real line 
were discussed
in \cite{FaddeevTakhtajan} using the inverse scattering method.
There is a difference in notations: 
our $\lambda_j$ differ from $\lambda_j$
of \cite{FaddeevTakhtajan} by a factor of $i$. 
The scattering data of the general solution includes a  
 discrete set of real (in our notations) $\lambda_j=\kappa_j$,
$\kappa_j\in {\bf R}$. Besides that, there is a discrete
set of complex conjugate pairs  $(\lambda_j,\overline{\lambda_j})$ with 
$\mbox{Im}\;\lambda_j\neq 0$ and also a continuous data
parametrized by a function $b(x)$ with $\overline{b}(x)=b(-x)$.
General solutions can be approximated by the rational solutions,
which have $b(x)=0$.  Therefore rational solutions depend
only on the discrete set of parameters $\kappa_j$ and 
$(\lambda_k,\overline{\lambda}_k)$.
It is useful to look at the asymptotic form of these rational
solutions in the infinite future, when $t=t_1+\tilde{t}_1=\infty$.
At $t=\infty$ the rational solutions split into well-separated
breathers (corresponding to $(\lambda_k,\overline{\lambda}_k)$)
and kinks (corresponding to $\kappa_j$). 
The energy of a breather can be made very small 
by putting $\lambda_k$ sufficiently close to the imaginary
axis (see Section \ref{sec:Breather}). This means that
one can continuously create a new breather from the vacuum.
In other words, creation of the new pair $(\lambda_k\overline{\lambda}_k)$
is a continuous operation; it changes a solution in the
continuous way. But the creation of a kink is not a continuous
operation. The creation of an odd
number of kinks would necessarily change the topological charge
of the solution. But even to create a pair of kink and anti-kink would require
a finite energy. This is our  justification for considering separately a sector
of solutions which do not have real $\lambda_j$. We will discuss
the action variable in this sector. 

\section{B\"acklund transformations and the ``hidden'' symmetry $U(1)_L$.}
\subsection{Construction of $U(1)_L$.}
\label{sec:Backlund}
Eq. (\ref{BacklundShift}) shows that the
B\"acklund transformations\footnote{more precisely, a particular solution
of the B\"acklund equations, defined as a series in $1/\mu$ or $\tilde{\mu}$} 
can be understood as a $\mu$-dependent
shift of times. We have two 1-parameter families of shifts
$B_{\mu}$ and $\tilde{B}_{\tilde{\mu}}$. We have 
$B_{\mu=\infty}={\bf 1}$ and $\tilde{B}_{\tilde{\mu}=0}={\bf 1}$.
It is not true that $B_{\mu}$ or $\tilde{B}_{\tilde{\mu}}$
is a one-parameter group of transformations, because
it is not true that $B_{\mu_1}B_{\mu_2}$ is equal to  $B_{\mu_3}$
with some $\mu_3$. 
\begin{figure}
\begin{center} 
\epsfxsize=3in {\epsfbox{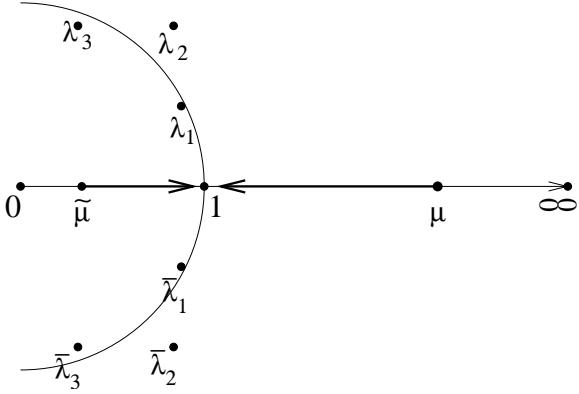}} 
\caption{The B\"acklund transformation $B_{\mu}$ can be
expanded in $1/\mu$ near $\mu=\infty$ and
$\tilde{B}_{\tilde{\mu}}$ can be expanded in $\tilde{\mu}$ near
$\tilde{\mu}=0$. 
We analytically continue $B$ and $\tilde{B}$ to $\mu=\tilde{\mu}=1$}
\end{center} 
\end{figure}
Both $B_{\mu}$ and $\tilde{B}_{\tilde{\mu}}$ preserve the
symplectic structure. Therefore we can discuss the
Hamiltonian vector fields $\xi_{\mu}$ and $\tilde{\xi}_{\tilde{\mu}}$
such that:
\begin{equation}
	e^{\xi_{\mu}}=B_{\mu},\;\;\;
	e^{\tilde{\xi}_{\tilde{\mu}}}=\tilde{B}_{\tilde{\mu}}
\end{equation}
\begin{figure}
\begin{center} 
\epsfxsize=3.7in {\epsfbox{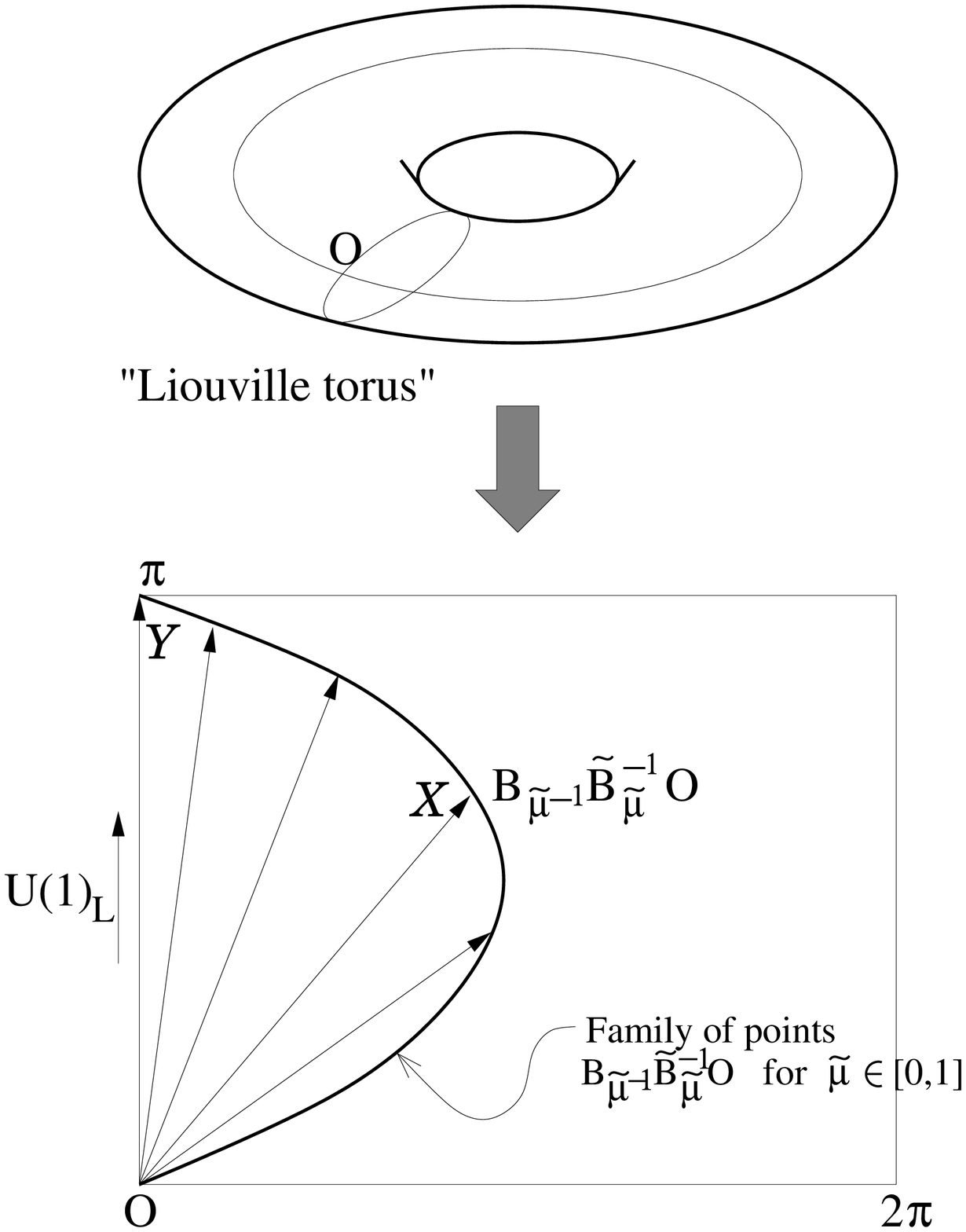}} 
\caption{The relation between the hidden symmetry $U(1)_L$ 
and the B\"acklund transformations. 
We pick a point $O$ on a ``Liouville torus'' (for a finite-gap
solution, this is an actual finite-dimensional torus).
The solid curve $OXY$ represents the 1-parameter family of points
$B_{\tilde{\mu}^{-1}}\tilde{B}_{\tilde{\mu}}^{-1}O$ 
parametrized by $\tilde{\mu}\in [0,1]$.
When $\tilde{\mu}=1$ the interval $OY$ (where 
$Y=B_{1}\tilde{B}_{1}^{-1}O$) is the b-cycle of the torus. 
The symmetry $U(1)_L$ acts by
shifts along this cycle. The arrow $OX$ represents the 1-parameter
family of points 
$\exp\left[t\log(B_{\tilde{\mu}^{-1}}\tilde{B}_{\tilde{\mu}}^{-1})
\right]O$ with $t\in [0,1]$. 
When $\tilde{\mu}\to 1$ the 1-parameter group 
$\exp\left[t\log(B_1\tilde{B}_1^{-1})\right]$ with $t\in [0,2]$ 
is $U(1)_L$.}
\end{center} 
\end{figure}
One could imagine an ambiguity in the definition of $\xi_{\mu}$
and $\tilde{\xi}_{\tilde{\mu}}$, but we have the continuous
families 
connecting $B_{\mu}$ to ${\bf 1}=B_{\infty}$ and 
$\tilde{B}_{\tilde{\mu}}$ to ${\bf 1}=\tilde{B}_0$.
The existence of these continuous families allows us to define
$\xi_{\mu}$ and $\tilde{\xi}_{\tilde{\mu}}$ unambiguously,
see Fig. 1. The formula is:
\begin{eqnarray}
	\xi_{\mu}=-\sum_{p=0}^{\infty} {\mu^{-2p-1}\over 2p+1}
	{\partial\over\partial t_{2p+1}}
	\label{XiMu}
	\\[5pt]
	\tilde{\xi}_{\tilde{\mu}}=\sum_{p=0}^{\infty}
	{\tilde{\mu}^{2p+1}\over 2p+1}
	{\partial\over\partial \tilde{t}_{2p+1}}
	\label{TildeXiTildeMu}
\end{eqnarray}
These vector fields act on the rational solutions through
the parameters $b_j$:
\begin{eqnarray}
&&	\xi_{\mu}.b_j={1\over 2}
	\log\left[{1-{\lambda_j/\mu}
	\over 
	1+{\lambda_j/\mu}}\right]
	b_j
	\\[5pt]
&&	\tilde{\xi}_{\tilde{\mu}}.b_j=
	{1\over 2}
	\log\left[{1-{\tilde{\mu}/\lambda_j}
	\over
	1+{\tilde{\mu}/\lambda_j}}\right]
	b_j
\end{eqnarray}
Let us consider the limit:
\begin{equation}\label{ActionVsTimes}
	\xi=\lim\limits_{\epsilon\to 0+}
	(\xi_{e^{\epsilon}}-\tilde{\xi}_{e^{-\epsilon}})
	=\lim_{\epsilon\to 0+}\sum_{p=0}^{\infty}
	\left[-{e^{-(2p+1)\epsilon}\over 2p+1}
	{\partial\over\partial t_{2p+1}}
	-{e^{-(2p+1)\epsilon}\over 2p+1}
	{\partial\over\partial \tilde{t}_{2p+1}}
	\right]
\end{equation}
We have:
\begin{equation}
	\xi.b_j=-{\pi i\over 2} \;\mbox{sign}(\mbox{Im}(\lambda_j))\;b_j
\end{equation}
We see that the trajectories of the  vector field $\xi$  are periodic:
\begin{equation}
	e^{2\xi}={\bf 1}
\end{equation}
Therefore $\xi$ is the Hamiltonian vector field of an 
{\em action variable}.
We denote $U(1)_L$ the corresponding hidden symmetry.
Notice that $e^{\xi}$ exchanges $\tau_+$ and $\tau_-$
and therefore maps $\phi\mapsto -\phi$:
\begin{equation}
	e^{\xi}=[\phi\mapsto -\phi]
\end{equation}
The corresponding symmetry of the classical string is 
$\vec{n}\mapsto -\vec{n}$.
We see that the discrete {\em geometric} 
${\bf Z}_2$-symmetry (the ``reflection'' $\phi\mapsto -\phi$) is related
to the continuous {\em hidden} symmetry $U(1)_L$ 
(generated by the higher Hamiltonians).
This example is of the same nature as the relation between
the anomalous dimension and the local charges discussed in
\cite{Anomalous}.

In the language of free fermions, $B_{\mu}$ corresponds to 
the creation of the free fermion $\psi(\mu)$ from the left vacuum,
and $\tilde{B}_{\tilde{\mu}}^{-1}$ 
to the creation of $\tilde{\psi}(\tilde{\mu})$
from the right vacuum. When $\mu\to\tilde{\mu}$, 
the leading term in the operator product expansion of
$\psi(\mu)\tilde{\psi}(\tilde{\mu})$ 
is a $c$-number, and it cancels between $\tau_+$ and $\tau_-$ in 
(\ref{PhiFromTau}). 
The shift of the
charge of the left and right Dirac vacua
leads to the exchange $\tau_+\leftrightarrow\tau_-$,
and therefore Eq. (\ref{PhiFromTau}) gives $\phi\mapsto -\phi$.

This construction essentially used the fact that 
$B_1\tilde{B}_1^{-1}=[\phi\mapsto -\phi]$.
In fact for any real $\mu=\tilde{\mu}$ we have
\begin{equation}\label{BP}
B_{\mu}\tilde{B}^{-1}_{\tilde{\mu}}|_{\tilde{\mu}=\mu}=
[\phi\mapsto -\phi]
\end{equation}
This can be understood directly from 
(\ref{LeftBacklundSmall}),
(\ref{LeftBacklundLarge}),
(\ref{RightBacklundSmall}),
(\ref{RightBacklundLarge}).
First of all we have to explain the meaning of the left hand
side of (\ref{BP}), because we defined $B_{\mu}$ only for large
$\mu$ as a series in $1/\mu$ and $\tilde{B}_{\tilde{\mu}}$
for small $\tilde{\mu}$ as a series in $\tilde{\mu}$.
Let us consider the null-surface limit (\ref{NSL}) and
construct $B_{\mu}$ and $\tilde{B}_{\tilde{\mu}}$ as a series
in $\epsilon$, where $\epsilon$ is the small parameter
of the null-surface perturbation theory defined in (\ref{NSL}).
In this perturbation theory we have 
$|\partial_{\sigma}\phi| << |\partial_{\tau}\phi|$ and $|\phi|<<1$.
The zeroth approximation  to (\ref{LeftBacklundSmall}),
(\ref{LeftBacklundLarge}) and (\ref{RightBacklundSmall}),
(\ref{RightBacklundLarge}) is:
\begin{eqnarray}\label{LeadingOrderBT}
	B_{\mu}\phi={1-\mu^{-2}\over 1+\mu^{-2}}\phi
	-{2\mu^{-1}\over 1+\mu^{-2}}\partial_{\tau}\phi
	\\[5pt]
	\label{LeadingOrderBTtilde}
	\tilde{B}_{\tilde{\mu}}\phi=
	{1-\tilde{\mu}^2\over 1+\tilde{\mu}^2}\phi
	-{2\tilde{\mu}\over 1+\tilde{\mu}^2}\partial_{\tau}\phi
\end{eqnarray}
We see that in the leading order
of the null-surface perturbation theory $B_{\mu}$ and
$\tilde{B}_{\tilde{\mu}}$ both depend on  $\mu$ and $\tilde{\mu}$
as rational functions. The higher orders are also 
rational functions of $\mu$ and $\tilde{\mu}$. 
Therefore in the null-surface perturbation theory $B_{\mu}$ 
and $\tilde{B}_{\tilde{\mu}}$ both have an unambiguous
analytic continuation to finite values of $\mu$ and $\tilde{\mu}$.
Therefore we can take $\mu=\tilde{\mu}$ and (\ref{BP})
follows from (\ref{LeftBacklundSmall}),
(\ref{LeftBacklundLarge}),
(\ref{RightBacklundSmall}),
(\ref{RightBacklundLarge}) 
and (\ref{LeadingOrderBT}),
(\ref{LeadingOrderBTtilde}). 
This means that the generator of $U(1)_L$ which we defined
in (\ref{ActionVsTimes}) as $\xi_1-\tilde{\xi}_1$ can be also
defined as $\xi_{\mu}-\tilde{\xi_{\mu}}$ for any real $\mu$.
For the rational solutions with all $\lambda_j$
having a nonzero imaginary part we have
\begin{equation}\label{LorentzInvariance}
	(\xi_{\mu}-\tilde{\xi}_{\mu})-
	(\xi_1-\tilde{\xi}_1)=0
\end{equation}
for any $\mu\in {\bf R}$.
Therefore $U(1)_L$ commutes with the Lorentz boosts which 
transform
$\phi(t_1,\tilde{t}_1)$ to $\phi(\mu t_1,\mu^{-1}\tilde{t}_1)$.

\subsection{Free field limit.}\label{sec:FreeFieldLimit}
In the limit $\phi\to 0$ the equations of motion become
\begin{equation}
	\partial_{t_1}\partial_{\tilde{t}_1}\phi=-4\phi
\end{equation}
And the left and right B\"acklund transformations become:
\begin{eqnarray}
	B_{\mu}.\phi={1-{1\over 2\mu}{\partial\over\partial t_1}
	\over 1+{1\over 2\mu}{\partial\over\partial t_1}}\phi
	\\[5pt]
	\tilde{B}_{\tilde{\mu}}.\phi=
	{1+{\tilde{\mu}\over 2}{\partial\over\partial \tilde{t}_1}\over
	1-{\tilde{\mu}\over 2}{\partial\over\partial \tilde{t}_1}}\phi
\end{eqnarray}
This means that in the free field limit:
\begin{eqnarray}
&&	{\partial\over\partial t_{2p+1}}\phi = 
	{1\over 2^{2p}} \left({\partial\over\partial t_1}\right)^{2p+1}\phi \\
&&	{\partial\over\partial \tcon_{2p+1}}\phi=
	{1\over 2^{2p}} 
	\left({\partial\over\partial \tilde{t}_1}\right)^{2p+1}\phi
\end{eqnarray}
The generator of $U(1)_L$ acts as follows:
\begin{eqnarray}
	\xi.\phi=
	\lim_{\epsilon\to 0+}\log\left[ 
	{(2-e^{-\epsilon}\partial_{t_1})
	 (2-e^{-\epsilon}\partial_{\tilde{t}_1})\over
	(2+e^{-\epsilon}\partial_{t_1})
	(2+e^{-\epsilon}\partial_{\tilde{t}_1})}
	 \right]\phi=
	 \nonumber
	 \\ =
	 \lim_{\epsilon\to 0+}\log\left[ 
	 {4\sinh\epsilon - \partial_{t_1}-\partial_{\tilde{t}_1} \over
	 4\sinh\epsilon + \partial_{t_1}+\partial_{\tilde{t}_1}}
	 \right]\phi
	 \label{XiOnFreeField}
 \end{eqnarray}
 The free field  $\phi$ has an oscillator expansion:
\begin{equation}
	\phi=\int{d k\over \sqrt{2\omega_k}}
	\left(\alpha_k e^{ik\sigma + i\omega_k\tau}+
	\overline{\alpha_k}e^{-ik\sigma-i\omega_k\tau}\right)
\end{equation}
where $\omega_k=\sqrt{4+k^2}$. Eq. (\ref{XiOnFreeField}) implies
that $U(1)_L$ is the oscillator number:
\begin{eqnarray}
&&	\xi. \alpha_k= \pi i\alpha_k \nonumber\\
&&	\xi. \overline{\alpha_k}= -\pi i\overline{\alpha_k}
\end{eqnarray}
This is in agreement with the results of \cite{PWL} and shows
that the $U(1)_L$ considered here is the same $U(1)_L$ as considered in
\cite{Notes,PWL}.

\subsection{Action of $U(1)_L$ on a breather.}\label{sec:Breather}
Consider $\lambda_1=\lambda=ie^{i\theta}|\lambda|$, 
$\lambda_2=\overline{\lambda}$,
$b_1=be^{i\varphi}$ and $b_2=b e^{-i\varphi}$  and
denote $e^{\kappa}=b^2 /\tan\theta$. We get
\begin{eqnarray}
	\tau_{\pm}&=&{2b^2\over\tan\theta} e^{-2t\sin\theta |\lambda|
	+2\tilde{t}\sin\theta |\lambda|^{-1}}\times\\
&&	\left[
	\cosh (2t|\lambda|\sin\theta -2\tilde{t}|\lambda|^{-2}\sin\theta
	-\kappa)\pm \right.\\
&&	\left. \pm i\tan\theta \cos (2t|\lambda|\cos\theta+2\tilde{t}
	|\lambda|^{-1}\cos\theta+2\varphi)\right]
	\nonumber
\end{eqnarray}
Therefore
\begin{equation}\label{Breather}
	\tan{\phi\over 2}=
	\tan\theta 
	{\cos (2t|\lambda|\cos\theta+2\tilde{t}
	|\lambda|^{-1}\cos\theta+2\varphi)\over
	\cosh (2t|\lambda|\sin\theta -2\tilde{t}|\lambda|^{-1}\sin\theta
	-\kappa)}
\end{equation}
Remember that $t_1={1\over 4}(\tau+\sigma)$
and $\tilde{t}_1={1\over 4}(\tau-\sigma)$.
The limit $\theta\to 0$ corresponds to a circular null-string.
Indeed, with $|\lambda|=1$ 
Eqs. (\ref{WithAbsValues}) and (\ref{Breather}) imply in this limit
 that at $\tau=0$ we have
$\int_{-\infty}^{\infty} d\sigma |\partial_{\sigma}\vec{n}|=2\pi$.

The generator of $U(1)_L$ acts on a breather by shifting the phase $\varphi$:
\begin{equation}\label{PhaseShift}
	\xi={\pi\over 2}
	\mbox{sign}(\cos\theta){\partial\over\partial\varphi}
\end{equation}
The general solution without kinks can be approximated by collections of
breathers. The $U(1)_L$ will shift the phases of all the breathers
by the same amount. 

We have seen in Section \ref{sec:FreeFieldLimit} that the generator of
$U(1)_L$ can be also understood as the nonlinear analogue  of
the oscillator number.
On the other hand, we can see from Eqs. (\ref{Breather}) and
(\ref{PhaseShift}) that in the null surface limit
\begin{equation}
	\xi\simeq {\pi\over 4} \left(
	{\partial\over\partial t}+{\partial\over\partial \tilde{t}}
	\right)+\ldots
\end{equation}
where dots denote the terms subleading in the null surface limit.
The leading term is the energy of the string, and the subleading
terms are the higher conserved charges. 
The fact that the energy is the oscillator number plus corrections
was observed
already in 
the work of H.J.~de Vega, A.L.~Larsen and N.~Sanchez 
\cite{deVegaLarsenSanchez}\footnote{I want
to thank A.~Tseytlin for bringing my attention to this work.} .

\subsection{The null-surface limit and the ``improved'' currents
of \cite{ArutyunovStaudacher,Engquist}.}\label{sec:Improved}
Eq. (\ref{Breather}) shows that in the null-surface limit 
the parameters $\lambda_j$ 
are localized in the vicinity of $\pm i$: 
\begin{equation}\label{NSLimitLambda}
\lambda_j=\pm i + O(\epsilon)
\end{equation}
The action of the higher Hamiltonians on
the parameters $b_j$ follows from (\ref{Tau}):
\begin{equation}
	{\partial\over\partial t_{2p+1}} b_j =\lambda_j^{2p+1} b_j
\end{equation}
Consider the following linear combination of the higher Hamiltonian
vector fields:
\begin{equation}
	\sum\limits_{p=0}^l {l!\over p!(l-p)!} 
	{\partial\over\partial t_{2p+1}}
	b_j= \lambda_j(\lambda_j-i)^l(\lambda_j+i)^l b_j \simeq 
	\epsilon^l b_j
\end{equation}
We see that the vector fields
\begin{equation}
	\Xi_l=\sum\limits_{p=0}^l 
	{l!\over p!(l-p)!} {\partial\over\partial t_{2p+1}}
\end{equation}
are generated by the ``improved'' currents; 
the vector field  $\Xi_l$ is of the
order $\epsilon^l$ in the null-surface perturbation theory.

The improved currents used in \cite{ArutyunovStaudacher,Engquist}
involve both left and right times. Let us introduce the improved 
Hamiltonian vector 
fields ${\cal G}_k$ which acts on the parameters $b_j$ in the following way:
\begin{equation}
	{\cal G}_k.b_j=\left(\lambda_j-{1\over \lambda_j}\right)
	\left[\left(\lambda_j+{1\over \lambda_j}\right)
	\right]^{2k} b_j
\end{equation}
These vector fields are local and improved, in the sense
that ${\cal G}_k\simeq \epsilon^{2k}$ in the null-surface
limit  (\ref{NSLimitLambda}). For example
${\cal G}_0={\partial\over\partial t_1}+
{\partial\over\partial \tilde{t}_1}$, 
${\cal G}_1={\partial\over\partial t_3}+
{\partial\over\partial t_1}+
{\partial\over\partial \tilde{t}_1}+
{\partial\over\partial \tilde{t}_3}$ and
${\cal G}_2={\partial\over\partial t_5}
+3{\partial\over\partial t_3}+2{\partial\over\partial t_1}+
2{\partial\over\partial \tilde{t}_1} +
3{\partial\over\partial \tilde{t}_3}+{\partial\over\partial\tilde{t}_5}$.

\vspace{5pt}
\noindent 
The Hamiltonian vector fields 
${\partial\over\partial t_{2p+1}}+{\partial\over\partial \tilde{t}_{2p+1}}$
can be expressed through the improved vector fields:
\begin{equation}\label{TimesVsImproved}
{\partial\over\partial t_{2p+1}}+{\partial\over\partial \tilde{t}_{2p+1}}
=
\sum\limits_{n=0}^{p} {2^{-2n}} U_{2p,2n} {\cal G}_n
\end{equation}
where $U_{2p,2n}$ are the coefficients of the Chebyshev polynomials
of the second kind:
\begin{equation}
	U_{2p}(x)=\sum\limits_{k=0}^p U_{2p,2k} x^{2k}=
	{(x+i\sqrt{1-x^2})^{2p+1}-(x-i\sqrt{1-x^2})^{2p+1}\over
	2i \sqrt{1-x^2}}
\end{equation}
In the null-surface perturbation theory ${\cal G}_k\simeq \epsilon^{2k}$.
On the other hand, for $\lambda$ sufficiently close to  $i$ 
or $-i$ we have 
\begin{eqnarray}
	{\pi i\over 2} \mbox{sign}(\mbox{Im}\;\lambda)=
	{\pi\over 2}{\left(\lambda-\lambda^{-1}\right)\over
	\sqrt{4-\left(\lambda+\lambda^{-1}\right)^2}}
	=\nonumber\\=
	{\pi\over 4}
	\sum\limits_{k=0}^{\infty} {(2k)!\over 2^{2k}(k!)^2}
	\left(\lambda-{1\over\lambda}\right)
	\left[{1\over 2}\left(\lambda+{1\over\lambda}\right)\right]^{2k}
\end{eqnarray}
This implies that 
\begin{equation}\label{PerturbativeExpansion}
	\xi = -{\pi\over 4}\sum\limits_{k=0}^{\infty}
	{(2k)!\over 2^{4k}(k!)^2}{\cal G}_k
\end{equation}
We see that the generator of $U(1)_L$ is indeed an infinite
sum of local conserved charges, with only finitely many 
terms participating at each order of the null-surface perturbation
theory.

Eq. (\ref{PerturbativeExpansion}) can also be obtained
from Eqs. (\ref{ActionVsTimes}) and (\ref{TimesVsImproved})
using the formula
\begin{equation}
	\lim_{\epsilon\to 0+}\sum\limits_{p=0}^{\infty}
	{e^{-2p\epsilon}\over 2p+1}
	U_{2p}(x)=
	{\pi\over 4}
	\left(1-x^2\right)^{-1/2}
\end{equation}

\section{Summary and discussion of $O(N)$ model.}\label{sec:Summary}
In this section we will summarize our construction
of the action variable for the sine-Gordon model 
and outline the analogous construction for the $O(N)$ 
sigma model.
\subsection{The construction of the action variable for the
sine-Gordon model.}
The sine-Gordon model has infinitely many local conserved charges,
which are in involution with each other. Therefore each solution
$\phi(\tau,\sigma)$
defines an infinite-dimensional ``invariant torus'' which is defined
as the orbit
of $\phi(\tau,\sigma)$ under the Hamiltonian flows generated
by the local conserved charges. This ``invariant torus'' consists
of the solutions $\phi(\tau,\sigma,t_3,\tilde{t}_3,t_5,\tilde{t}_5,\ldots)$
where $t_{2n+1}$ and $\tilde{t}_{2n+1}$ are the ``higher times''
defined so that $\partial\over\partial t_{2n+1}$ and
$\partial\over\partial \tilde{t}_{2n+1}$ are the Hamiltonian
vector fields generated by the higher Hamiltonians.
We denote $t_1={1\over 4}(\tau+\sigma)$ and 
$\tilde{t}_1={1\over 4}(\tau-\sigma)$.

The B\"acklund transformations $B_{\mu}$ and $\tilde{B}_{\tilde{\mu}}$
depend on the parameters $\mu$ and $\tilde{\mu}$. They can be understood as
the shifts of the higher times;  $B_{\mu}$ is the shift
$t_{2n+1}\to t_{2n+1}-{\mu^{-2n-1}\over 2n+1}$
and $\tilde{B}_{\tilde{\mu}}$ is the shift
$\tilde{t}_{2n+1}\to \tilde{t}_{2n+1}+{\tilde{\mu}^{2n+1}\over 2n+1}$.
We have:
\begin{equation}\label{LeftBacklund}
	\begin{array}{l}
	{\partial\over\partial t_1}(B_{\mu}\phi+\phi)
	=-2\mu\sin(B_{\mu}\phi-\phi)\\
	{\partial\over\partial \tilde{t}_1}
	(B_{\mu}\phi-\phi)=
	{2\over\mu}\sin(B_{\mu}\phi+\phi)
\end{array}
\end{equation}
\begin{equation}\label{RightBacklund}
	\begin{array}{l}
	{\partial\over\partial \tilde{t}_1}
	(\tilde{B}_{\tilde{\mu}}\phi+\phi)=
	{2\over\tilde{\mu}}\sin(\tilde{B}_{\tilde{\mu}}\phi-\phi)
\\
	{\partial\over\partial t_1}
	(\tilde{B}_{\tilde{\mu}}\phi-\phi)=
	-2\tilde{\mu}\sin(\tilde{B}_{\tilde{\mu}}\phi+\phi)
\end{array}
\end{equation}
We define $B_{\mu}$ as a power series in $1/\mu$ and 
$\tilde{B}_{\tilde{\mu}}$ as a power series in $\tilde{\mu}$.
Then we define the ``logarithm'' of the B\"acklund transformation.
For each $\mu$ (large) and $\tilde{\mu}$  (small) 
the Hamiltonian vector field $\xi_{\mu,\tilde{\mu}}$ 
is generated by a linear combination of the
local Hamiltonians, such that
$e^{\xi_{\mu,\tilde{\mu}}}=B_{\mu}\tilde{B}^{-1}_{\tilde{\mu}}$.
The explicit formula is $\xi_{\mu,\tilde{\mu}}=
-\sum\left[{\mu^{-2p-1}\over 2p+1}{\partial\over\partial t_{2p+1}}
+{\tilde{\mu}^{2p+1}\over 2p+1}{\partial\over\partial \tilde{t}_{2p+1}}
\right]$.
We defined $B_{\mu}$ perturbatively around $\mu=\infty$
and $\tilde{B}_{\tilde{\mu}}$ perturbatively around
$\tilde{\mu}=0$. But we have seen that (at least on 
the rational solutions, which form a dense set) there
is a well-defined limit
$\xi=\lim\limits_{\mu\to 1,\tilde{\mu}\to 1}
\xi_{\mu,\tilde{\mu}}$.
Eqs. (\ref{LeftBacklund}) and (\ref{RightBacklund}) suggest
that $B_1\tilde{B}_1^{-1}$ is the transformation bringing
$\phi$ to $-\phi$, and we have seen that this is indeed the
case for rational solutions and in the null-surface perturbation
theory. 
Therefore $e^{2\xi}$ is the identical transformation,
which means that $\xi$ is generated by an  action variable. 

\subsection{The $O(N)$ sigma-model.}
The same arguments can be applied to the $O(N)$ sigma model.
The B\"acklund transformations are defined by the
same equations (\ref{LeftBacklundVector}) and 
(\ref{RightBacklundVector}) as for $N=3$, but 
instead of a 3-dimensional vector $\vec{n}(\tau,\sigma)$
we have an $N$-dimensional vector $X$:
\begin{equation}\label{LBX}
	\begin{array}{l}
	\partial_{\tilde{t}_1}(B_{\mu}X-X)=
	-{1\over 2}(1+\mu^{-2})(B_{\mu}X,\partial_{\tilde{t}_1}X)
	(B_{\mu}X+X)
	\\[5pt]
	\partial_{t_1}(B_{\mu}X+X)=
	{1\over 2}(1+\mu^2)(B_{\mu}X,\partial_{t_1}X)
	(B_{\mu}X-X)
\end{array}
\end{equation}
\begin{equation}\label{RBX}
	\begin{array}{l}
\partial_{t_1}(\tbmu X -X)=
	-{1\over 2}(1+\tmu^2)(\tbmu X,\partial_{t_1}X)
	(\tbmu X+X)
	\\[5pt]
	\partial_{\tilde{t}_1}(\tbmu X+X)=
	{1\over 2}(1+\tmu^{-2})(\tbmu X,\partial_{\tilde{t}_1}X)
	(\tbmu X-X)
\end{array}
\end{equation}
We conjecture that 
the B\"acklund transformations correspond to the shift of times
if we define them perturbatively as  series in $1/\mu$ and $\tilde{\mu}$.
(We do not know a proof of this fact for the $O(N)$ model.)
But we can also define the B\"acklund transformations perturbatively
using the null-surface perturbation theory. In the null-surface
perturbation theory the small parameter is $1/|p|=1/|\partial_{\tau}X|$
and $\partial_{\sigma}X$ remains finite.
In the limit $|p|\to\infty$ we observe that (\ref{LBX}) and (\ref{RBX})
are solved by:
\begin{eqnarray}\label{FirstApproximationLeft}
&&	B_{\mu}X={1-\mu^{-2}\over 1+\mu^{-2}}X-
	{2\mu^{-1}\over 1+\mu^{-2}}
	{\partial_{\tau}X\over |\partial_{\tau}X|}\\[5pt]
&&	\tilde{B}_{\tilde{\mu}}X={1-\tilde{\mu}^2\over 1+\tilde{\mu}^2}X+
	{2\tilde{\mu}\over 1+\tilde{\mu}^2}
	{\partial_{\tau}X\over |\partial_{\tau}X|}
	\label{FirstApproximationRight}
\end{eqnarray}
The corrections to (\ref{FirstApproximationLeft}),
(\ref{FirstApproximationRight})
by the higher powers of $1/|\partial_{\tau}X|$ involve
higher derivatives in $\tau$ and $\sigma$
and depend on $\mu$ as rational functions.
When $\mu$ is large we can expand these corrections 
in powers of $\mu^{-1}$.
Therefore the definition of the B\"acklund transformation $B_{\mu}$
as a power series in $1/|\partial_{\tau}X|$ agrees with the usual 
definition as a power series in $\mu^{-1}$, but does not require
$\mu$ to be large.

As we did for the sine-Gordon model, we can define $\xi_{\mu}$ as
the Hamiltonian vector field
$\log B_{\mu}$ and $\tilde{\xi}_{\tilde{\mu}}=\log B_{\tilde{\mu}}$.
Now we want to put $\mu=\tilde{\mu}$. There is a potential problem
here, because $B_{\mu}$ was defined as a series in $1/\mu$
and $\tilde{B}_{\tilde{\mu}}$ as a series in $\tilde{\mu}$.
But as we discussed, we can also define $B_{\mu}$ and $\tilde{B}_{\tilde{\mu}}$
in the null-surface perturbation theory using $1/|\partial_{\tau}X|$
as a small parameter. Then there is no problem 
doing the analytical continuation to $\mu=\tilde{\mu}$,
because at every order of the $1/|\partial_{\tau}X|$ perturbation theory
$B_{\mu}$ is a rational function of $\mu$.
(For example, the zeroth order is given by (\ref{FirstApproximationLeft}).)
Equations (\ref{LBX}), (\ref{RBX}), (\ref{FirstApproximationLeft})
and (\ref{FirstApproximationRight}) imply that
for $\mu=\tilde{\mu}$ we have $B_{\mu}\tilde{B}^{-1}_{\mu}X=-X$.
Therefore
$$e^{\xi_{\mu}-\tilde{\xi}_{\mu}}=[X\mapsto -X]$$ 
This implies that the Hamiltonian vector field
$\xi_{\mu}-\tilde{\xi}_{\mu}$ is generated by an action variable.
This vector field is independent of the choice of $\mu$ (in the
perturbation theory in $1/|\partial_{\tau}X|$)
because of the uniqueness of the action variable. 
For the explicit calculation it would be convenient
to choose $\mu=1$, because with this choice it is 
manifest that the action variable is a combination
of the local charges which is left-right symmetric.
The vector field 
${\partial\over\partial{\mu}}(\xi_{\mu}-\tilde{\xi}_{\mu})$
is zero in the perturbation theory because of the
relations between the left and right charges 
discussed in Section 3 of \cite{PWL}.

{\em Note in the revised version:} see \cite{Baecklund}
for the discussion of the $O(N)$ model.

\section*{Acknowledgments}
I want to thank N.~Beisert and V.~Kazakov for the
correspondence and explanations 
 of \cite{BKS,BKSZ}, and A.~Tseytlin for discussions 
of the fast moving strings.
This research was supported by the Sherman Fairchild 
Fellowship and in part
by the RFBR Grant No.  03-02-17373 and in part by the 
Russian Grant for the support of the scientific schools
NSh-1999.2003.2.


\begin{thebibliography}{10}
\bibitem{FT02}{S. Frolov, A.A. Tseytlin, 
"Semiclassical quantization of rotating
 superstring in $AdS_5 \times S^5$", JHEP 
{\bf 0206} (2002) 007, hep-th/0204226.}
\bibitem{Tseytlin}{A.A.~Tseytlin, "Semiclassical quantization of superstrings:
$AdS_5\times S^5$ and beyond", Int. J. Mod. Phys. {\bf A18} (2003) 981,
hep-th/0209116.}
\bibitem{Russo}{J.G.~Russo, "Anomalous dimensions in gauge theories from
rotating strings in $AdS_5\times S^5$," JHEP {\bf 0206} (2002) 038, 
hep-th/0205244. }
\bibitem{MinahanZarembo}{J. A. Minahan, K. Zarembo,
	``The Bethe-Ansatz for N=4 Super Yang-Mills,''
	JHEP 0303 (2003) 013, hep-th/0212208.}
\bibitem{FT03}{S. Frolov, A.A. Tseytlin,
	``Multi-spin string solutions in $AdS_5$ x $S^5$,''
	Nucl.Phys. {\bf B668} (2003) 77-110,
	hep-th/0304255.}
\bibitem{FTQ}{S.~Frolov, A.~A.~Tseytlin, ``Quantizing three-spin
	string solution in $AdS_5\times S^5$,''
	JHEP {\bf 0307} 016 (2003), 
	hep-th/0306130.}
\bibitem{Kruczenski}{M.~Kruczenski, 
``Spin chains and string theory'', hep-th/0311203.}
\bibitem{KRT}{M. Kruczenski, A.V. Ryzhov, A.A. Tseytlin,
	``Large spin limit of $AdS_5$ x $S^5$ string theory 
	and low energy expansion of ferromagnetic spin chains'',
	Nucl.Phys. {\bf B692} (2004) 3-49, hep-th/0403120.} 
\bibitem{MandalSuryanarayanaWadia}{G.~Mandal, N.V.~Suryanarayana,
	S.R.~Wadia, 
	``Aspects of Semiclassical
	Strings in AdS$_5$'', Phys.Lett. {\bf B543} (2002) 81,
	hep-th/0206103.}
\bibitem{BPR}{I.~Bena, J.~Polchinski, R.~Roiban,
``Hidden Symmetries of the $AdS_5 \times S^5$ Superstring'',
Phys.Rev. {\bf D69} (2004) 046002, hep-th/0305116.}
\bibitem{MartinWolf}{M.~Wolf, ``On Hidden Symmetries of a Super
	Gauge Theory and Twistor String Theory'',
	hep-th/0412163.}
\bibitem{DNW1}{L. Dolan, C.R. Nappi, E. Witten,
	``A Relation Between Approaches to Integrability in Superconformal 
	Yang-Mills Theory'', JHEP 0310 (2003) 017, hep-th/0308089.}
\bibitem{DNW2}{L. Dolan, C.R. Nappi, E. Witten,
	`` Yangian Symmetry in D=4 Superconformal Yang-Mills Theory'',
	hep-th/0401243.}
\bibitem{DolanNappi}{L. Dolan, C.R. Nappi,
	``Spin Models and Superconformal Yang-Mills Theory'',
	hep-th/0411020.}
\bibitem{Anomalous}{A.~Mikhailov, ``Anomalous dimension and local
	charges'', hep-th/0411178.}
\bibitem{Pohlmeyer}{K.~Pohlmeyer, ``Integrable Hamiltonian
	 Systems and Interactions through Quadratic Constraints'',
	 Comm. Math. Phys. {\bf 46} (1976) 207-221.}
\bibitem{PWL}{A.~Mikhailov, ``Plane wave limit of local conserved
	charges'', hep-th/0502097.}
\bibitem{Notes}{A.~Mikhailov, ``Notes on fast moving strings'',
	hep-th/0409040.}
\bibitem{ArutyunovStaudacher}{ G. Arutyunov, M. Staudacher,
	 ``Matching Higher Conserved Charges for Strings and Spins'',
JHEP 0403 (2004) 004, hep-th/0310182}
\bibitem{Engquist}{J.~Engquist,
	``Higher Conserved Charges and Integrability 
	for Spinning Strings in $AdS_5$ x $S^5$'',
	JHEP 0404 (2004) 002, hep-th/0402092.}
\bibitem{KT}{M. Kruczenski, A. Tseytlin,
	``Semiclassical relativistic strings in 
	$S^5$ and long coherent operators in N=4 SYM theory'',
	hep-th/0406189.}
\bibitem{dVGN}{H.J.~De~Vega, A.~Nicolaidis, ``Strings in strong
	gravitational fields'', Phys. Lett. {\bf B295} 214-218;
	H. J. de Vega, I. Giannakis, A. Nicolaidis,
	``String Quantization in Curved Spacetimes: Null String Approach'',
	Mod.Phys.Lett. A10 (1995) 2479-2484, hep-th/9412081.}
\bibitem{SpeedingStrings}{A.~Mikhailov, ``Speeding Strings'',
 JHEP 0312 (2003) 058, hep-th/0311019.}
\bibitem{ArutyunovFrolov0411}{G.~Arutyunov, S.~Frolov,
	``Integrable Hamiltonian for Classical Strings on 
	$AdS_5\times S^5$'', JHEP 0502 (2005) 059,
	hep-th/0411089.}
\bibitem{BKSZ}{N. Beisert, V. A. Kazakov, K. Sakai, K. Zarembo,
	``The Algebraic Curve of Classical Superstrings 
	on $AdS_5\times S^5$'',
	hep-th/0502226.}
\bibitem{AAT}{L. F. Alday, G. Arutyunov, A. A. Tseytlin,
	``On Integrability of Classical SuperStrings in 
	$AdS_5\times S^5$'', hep-th/0502240.}
\bibitem{Minahan}{J.~Minahan, 
	``Higher Loops Beyond the SU(2) Sector'', JHEP 0410 (2004) 053,
	hep-th/0405243; ``The SU(2) sector in AdS/CFT'', hep-th/0503143.}
\bibitem{BKS}{N. Beisert, V. A. Kazakov, K. Sakai,
	``Algebraic Curve for the SO(6) sector of AdS/CFT'',
	hep-th/0410253.}
\bibitem{KMMZ}{V.A.Kazakov, A.Marshakov, J.A.Minahan, K.Zarembo,
	``Classical/quantum integrability in AdS/CFT'',
	JHEP 0405 (2004) 024, hep-th/0402207.}
\bibitem{KazakovZarembo}{ V.A. Kazakov, K. Zarembo,
	``Classical/quantum integrability in non-compact sector of AdS/CFT'',
	hep-th/0410105.}
\bibitem{SchaferNameki}{S.~Schafer-Nameki,
	``The Algebraic Curve of 1-loop Planar N=4 SYM'',
	hep-th/0412254.}
\bibitem{Nonlocal}{A.~Mikhailov, ``A nonlocal Poisson bracket
	of the sine-Gordon model'', hep-th/0511069.}
\bibitem{BabelonBernardAffine}{O.~Babelon, D.~Bernard,
	``Affine Solitons: A Relation Between Tau Functions, Dressing
	and B\"acklund Transformations'', hep-th/9206002.}
\bibitem{FaddeevTakhtajan}{L.D.~Faddeev, L.A.~Takhtajan,
	``Hamiltonian Methods in the Theory of Solitons'',
	Springer-Verlag, Berlin, 1987.}
\bibitem{KMMMZ}{S.~Kharchev, A.~Marshakov, A.~Mironov, A.~Morozov, 
	A.~Zabrodin,
	``Towards unified theory of $2d$ gravity'', 
	Nucl. Phys. {\bf B380} (1992) 181-240, hep-th/9201013.}
\bibitem{NeveuPapanicolaou}{A. Neveu, N. Papanicolaou,
	``Integrability of the classical scalar and symmetric 
	scalar - pseudoscalar contact Fermi interactions in two-dimensions'',
	Commun.Math.Phys.58:31,1978.}
\bibitem{deVegaLarsenSanchez}{H.J.~de~Vega, A.L.~Larsen, N.~Sanchez,
	``Semi-Classical Quantization of Circular Strings 
	in De Sitter and Anti De Sitter Spacetimes'',
	Phys.Rev. D51 (1995) 6917-6928, hep-th/9410219.}
\bibitem{Baecklund}{A.~Mikhailov, ``B\"acklund transformations, 
	energy shift and the plane wave limit'',
	hep-th/0507261.}
\end{thebibliography}
\end{document}